\documentclass[aps,pre,twocolumn,floatfix]{revtex4}

\usepackage[utf8]{inputenc}
\usepackage{amsmath}
\usepackage{graphicx}
\usepackage{amssymb}
\usepackage{epsfig}
\usepackage{natbib}
\usepackage{algorithm}				

\def\bq{\begin{equation}}
\def\eq{\end{equation}}

\begin{document}

\title{L\'evy ratchets in the spatially tempered \\fractional Fokker-Planck equation}

\author{A. Kullberg}
\affiliation{Department of Physics and Astronomy\\ University of California, Los Angeles \\ Los Angeles, California 90095}
\author{D. del-Castillo-Negrete}
\email{delcastillod@ornl.gov}
\affiliation{Oak Ridge National Laboratory \\ Oak Ridge TN, 37831-8071}

\begin{abstract}
L\'evy ratchets are minimal models of fluctuation-driven transport in the presence of  L\'evy noise and 
periodic external potentials with broken spatial symmetry. L\'evy noise can drive the system out-of thermodynamics equilibrium and previous works have shown that in this case, a net ratchet current can appear  even in the absence of time dependent perturbations, external tilting forces,  or a bias in the noise.  The majority of  studies on the interaction of   L\'evy noise with external potentials have assumed $\alpha$-stable L\'evy statistics in the Langevin description, which in the continuum limit corresponds to the fractional Fokker-Planck equation. However, the 
divergence of the low order moments is a potential drawback of $\alpha$-stable distributions because, in applications, the moments represent physical quantities. For example, for $\alpha <1$, the current (first moment), $J$, in $\alpha$-stable L\'evy ratchets is unbounded.  To overcome this limitation, in this paper we study ratchet transport using truncated L\'evy distributions which in the continuum limit correspond to the spatially tempered fractional Fokker-Planck equation. The main object of study is the dependence of the ratchet current on the level of tempering, $\lambda$. For finite tempering, $\lambda \neq 0$,  the statistics ultimately converges (although very slowly) to Gaussian diffusion in the absence of a potential. However,
it is shown here that in the presence of a ratchet potential a finite non-equilibrium current persists asymptotically for any finite value of $\lambda$.  The current is observed to converge exponentially in time to the steady state value. 
The steady state current exhibits algebraically decay with $\lambda$, as $J\sim \lambda^{-\zeta}$, for $\alpha \geq 1.75$. However,   for $\alpha \leq 1.5$, the  steady state current  decays exponentially with $\lambda$,  as $J \sim e^{-\xi \lambda}$. In the presence of a bias in the  L\'evy noise, it is shown that the 
tempering can lead to a current reversal. A detailed numerical study is presented on the dependence of the current on $\lambda$ and the physical parameters of the system. 

\end{abstract}

\maketitle

\section{Introduction}

%
%

The problem of fluctuation driven transport has a long and interesting history, with early discussions dating back to work by Smoluchowsky in 1912. For a review,  see 
Ref.~\cite{reimann} and references therein. Of particular interest are periodic potentials with broken spatial symmetry, known as ratchet potentials, see for example Fig.~\ref{fig_ratch_pot}.  In this case, particles in the potential wells experience an asymmetric force and 
one might naively expect that in the presence of thermal equilibrium  fluctuations particles will, on the average, drift down-hill in the direction of the weaker force, giving rise to a net current. However, it has long been recognized that this naive picture is incorrect since net work cannot be extracted out of equilibrium thermal fluctuations. However, in the presence of non-equilibrium perturbations the situation is quite different and  net currents can be generated. For example, in the case of pulsating ratchets a current can be generated by driving the system out of thermal equilibrium by adding a time variation to the amplitude of the potential. 
Considerable amount of work has thus been devoted to construct ``minimal" models of ratchet transport which involve various far from equilibrium mechanisms.  However, most of these studies have focused on models in which the fluctuations satisfy Gaussian statistics. 
\\

%
%
The early work reported in Refs.~\cite{dcn_preprint,dcn_physA} 
originally proposed a minimal model for L\'evy ratchets consisting  of a time-independent ratchet potential in the presence of  uncorrelated L\'evy noise. It was shown that, 
even in the absence of an external tilting force or time dependence in the potential, or a bias in the noise, the L\'evy flights drive the system out of thermodynamic equilibrium and generate an up-hill current (i.e., a current in the direction of the steeper side of the asymmetric potential). Following this, Ref.~\cite{dybiec_2008} studied the L\'evy ratchet problem 
and proposed robust probability measures of directionality of transport. 
In Ref.~\cite{dybiec_2008_b}  the influence of periodically modulated L\'evy noise asymmetry was studied.
Beyond the study of ratchet transport, the study of the interaction of L\'evy noise with external potentials has been a topic of considerable recent interest.  Some examples include the study harmonic  \cite{jespersen1999,chechkin_2000} and non-harmonic \cite{chechkin_2002, samorodnitsky_2003,dubkov_2007} potentials, and the study of the 
Kramers' problem in
Refs.~\cite{ditlevsen_1999,chechkin_2005,dybiec_2007,imkeller_2006}.
\\
%
%

L\'evy statistics has been used to model a wide range of problems involving probability density functions (pdfs) with slowly decaying tails. However, an often overlooked issue is the differences that typically exist between the strict mathematical properties of L\'evy distributions and the pdfs found in practical problems
involving experiments or numerical simulations. 
For example, whereas   the second and higher moments of $\alpha$-stable L\'evy distribution with $1<\alpha<2$ are formally infinite, the moments of pdfs of practical interest are finite. 
One of the reasons why pdfs of practical interest have finite moments is that they typically have finite support because the statistical samples upon which they are constructed are finite. For example, pdfs with algebraic decaying tails have been observed in the description of particle displacements in turbulent transport \cite{del_castillo_2005}. These numerically determined pdfs have finite moments because their support is limited by the largest possible displacement which is finite. 
Another key issue is that in applications, the moments represent physical quantities that cannot be infinite. 
This issue is particularly delicate  in the study of L\'evy flights in external potentials where  $\alpha$-stable  L\'evy noise is added to the Langevin stochastic equation for the particle velocity. In this case, the divergence of the second moment can lead in principle to a divergence in the kinetic energy. 
In practical terms this problem might not be disastrous because any numerical realization of the numerical model will have finite moments.  Nevertheless this shortcoming of the L\'evy $\alpha$-stable Langevin model, which is also present in the corresponding $\alpha$-stable fractional Fokker-Planck model, has to be resolved.  In this paper we propose to use the truncated fractional diffusion operators introduced in Ref.~\cite{cartea_del_castillo_2007} to address this problem.

%
%
%
%

Truncated L\'evy distributions distributions were originally introduced in  Ref.\cite{mantegna,kopone} as a simple prescription to guarantee the finiteness of the second moment. Reference~\cite{rosinski}   considered a general class of multivariate tempered stable L\'evy processes and established their parametrization and probabilistic representations. 
General L\'evy processes, and in particular tempered stable processes, were incorporated in the continuous time random walk model and in macroscopic transport models in Ref.~\cite{cartea_del_castillo_2007}. This reference introduced the notion of truncated fractional derivatives in the construction of non-diffusive transport models driven by truncated L\'evy flights. 
The role of tempered L\'evy distribution in the super-diffusive propagation of fronts in reaction-diffusion systems has been recently studied in Ref.~\cite{del_castillo_2009}. 
%
%

The goal of this paper is to study the role of truncation on fluctuation driven transport with L\'evy noise. 
%
%
Our approach is based on the  numerical solution of the spatially tempered Fractional Fokker Planck equation, obtained  
by replacing the diffusion operator by a truncated fractional diffusion operator. 
We use two complementary numerical methods. One is based on the finite difference, Grunwald-Letnikov discretization of the 
truncated fractional derivatives that allows the computation of the space-time evolution of the pdf in a finite size domain. The second method is a Fourier based spectral method that allows the computation of the reduced pdf in a periodic domain as well as the time dependent and asymptotic steady state current. 
The main object of study is the dependence of the steady state current on the level of truncation, $\lambda$, the stability index $\alpha$, and the noise asymmetry $\theta$ as well as the asymmetry of the potential.

%
%
The rest of the paper is organized as follows. In the next section we discuss the setting of the  L\'evy ratchet problem and introduce the spatially tempered Fractional Fokker Planck equation. Section~III presents the numerical results obtained from the solution of the time dependent and the steady state tempered Fractional Fokker Planck equation. Section~IV contains the conclusions.

\section{Spatially tempered fractional Fokker-Planck equation}


The most natural formulation of the problem of fluctuation-driven transport in the presence of an external potential is based on the Langevin approach. The starting point is an ensemble of particles with coordinates $x(t)$  which in the over-damped case evolve according to the stochastic differential equation 
\bq
\label{langevin}
dx = -\left( \partial_x V \right) dt + d \eta 
\eq
where for simplicity, and without loss of generality, we have assumed a one-dimensional domain. $V(x)$ denotes the external potential, and $d \eta$ is a stochastic processes modeling the fluctuations.  In the case of uncorrelated, Gaussian fluctuations,
 $d \eta=\sqrt{2 \chi} dW$, where $dW$  is a stochastic Wiener process, with $\langle dW \rangle=0$ and $\langle dW^2 \rangle= dt$.    As it is well-known, if the individual particles evolve according to the Langevin equation (\ref{langevin})  with uncorrelated Gaussian noise, then the probability density function (pdf) satisfies the Fokker-Plank equation
\bq
\label{fp_eq}
\partial_t P =  \partial_x \left[ P \partial_x V \right] + \chi \partial_x^2 P \, . 
\eq

\begin{figure}
\includegraphics[scale=0.4]{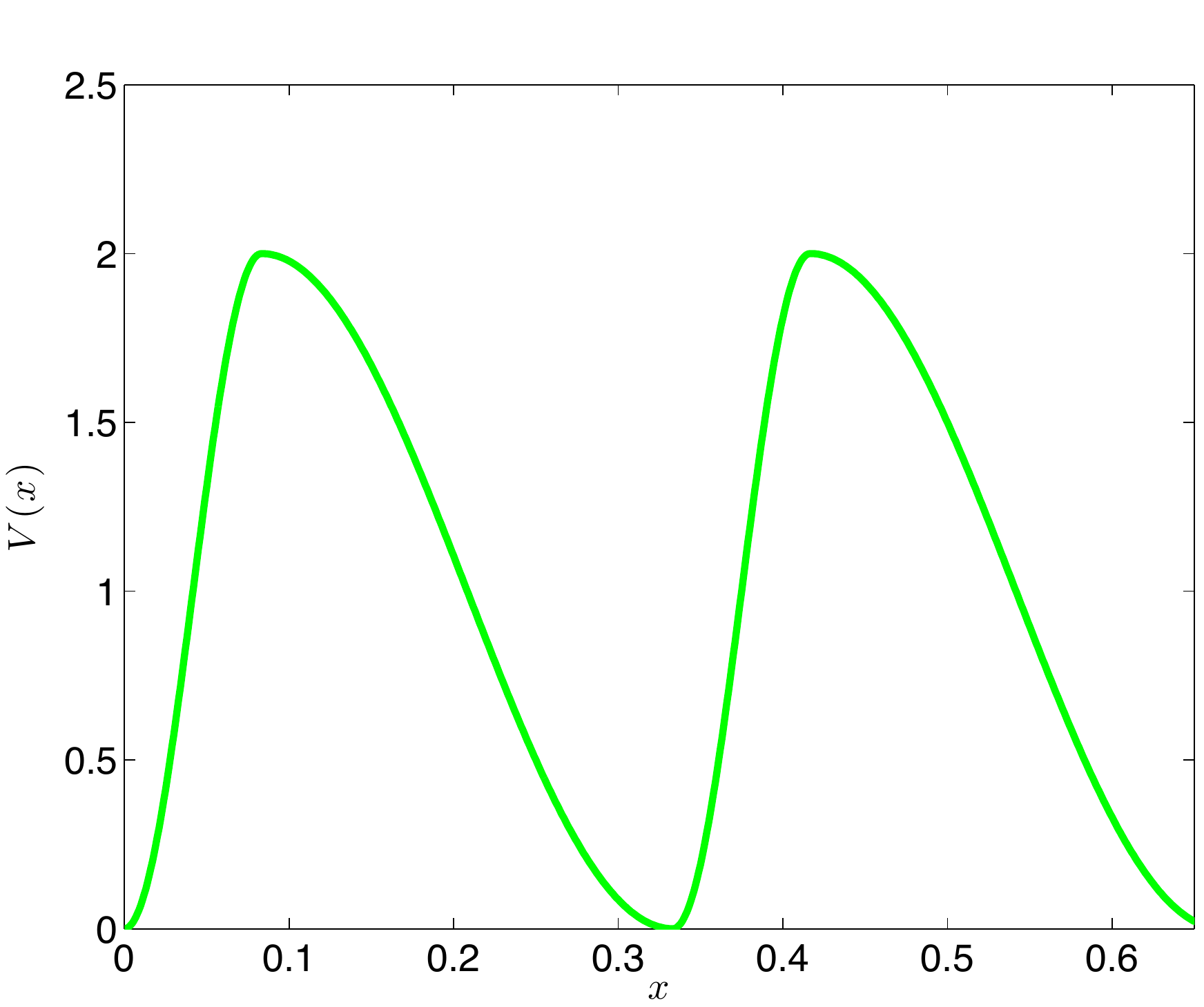} 
\caption{
\label{fig_ratch_pot} 
Ratchet potential in Eq.~(\ref{potential}) for $A=-0.5$ and $V_0=1$.
}
\end{figure}

Of particular relevance to the present paper is the study of 
periodic potentials with broken spatial symmetry. These, potentials, also called ``ratchets",  satisfy $V(x)=V(x+L)$ where  $L$ is the length of the period but they lack reflection symmetry. A paradigmatic  example is  
 $V=V_0[ \sin(2 \pi x/L) +  0.25 \sin(4 \pi x/L)]$. For an easier control of the degree of spatial symmetry, following
 \cite{dcn_physA}, here we use 
 \bq
\label{potential}
V=V_0\left\{ 
\begin{array}{ll}
1- \cos\left[ \pi x/a_1 \right] & \mbox{if $0 \leq x < a_1$} \\
1+ \cos\left[ \pi (x-a_1)/a_2 \right] & \mbox{if $a_1 \leq x < L$} \, ,\\
\end {array}
\right.
\eq
where $V_0$ is the  amplitude, $L=a_1+a_2$ is the period,  and $A=(a_1-a_2)/L$
is the asymmetry parameter. In all the calculations presented here, $V_0=L=1$.  In the study of ratchets it is customary to add an external constant ``tilting" force $F$ to the potential, and to consider the effective potential $V_{eff}=V-Fx$. However, unless mentioned otherwise, the calculations presented here will for the most part assume $F=0$. 
Figure~\ref{fig_ratch_pot}  shows a plot of the ratchet potential in Eq.~(\ref{potential}) with
$a_1=1/4$, $a_2=3/4$, $V_0=1$, and $L=1$, that correspond to $A=-0.5$. 

The conditions for the existence of ratchet currents in the presence of diffusive transport have been extensively studied and are well-understood, see for example Ref.~\cite{reimann} and references therein. However this is not the case for fluctuation-driven transport in the presence of non-diffusive transport.  To study the role non-diffusive transport in fluctuation-driven ratchet transport, Ref.~\cite{dcn_physA} considered the 
Langevin model but with $d \eta$ corresponding to a L\'evy process.
In this case, the continuum limit of the Langevin equation leads to the   $\alpha$-stable fractional Fokker-Planck equation
\bq
\label{eq_23}
\partial_t P = \partial_x \left[ P \partial_x V \right] + \chi  \partial_x^\alpha P \, .
\eq
where
\bq
\label{alpha_operator}
\partial_x^\alpha P =  l _{-\infty}D_x^\alpha P + 
r _xD_\infty^\alpha  P \, .
\eq
The operators on the right hand side of Eq.(\ref{alpha_operator}) are the left and right 
Riemann-Liouville fractional derivatives which are integro-differential operators defined in Fourier space according to
\cite{podlu_1999,samko_1993}
\bq
\label{fd_rl}
{\cal F} \left[ _{-\infty}D_x^\alpha P \right]=\left(-i k \right)^\alpha \tilde{P}\, , \qquad
{\cal F} \left[ _{x}D_{\infty}^\alpha P \right]=\left(i k \right)^\alpha \tilde{P}\, , 
\eq
where ${\cal F}\left[ f \right](k)=\tilde{f}=\int f(x) e^{ikx} dx$.
The  factors
\bq
l= - \frac{(1-\theta)}{2 \cos(\alpha \pi /2)}\, , \qquad r= - \frac{(1+\theta)}{2 \cos(\alpha \pi /2)} \, ,
\eq
determine the relative
weight of the left and the right fractional derivatives in terms of the parameter $\theta$. The order
of the fractional operators $\alpha$, the asymmetry parameter $\theta$,  and the diffusivity $\chi$
correspond to  the index, the skewness,  and the scale factor of the characteristic function of the L{\' e}vy noise in the Langevin formulation. 


From the statistical mechanics point of view, the difference between the transport operators in  Eqs.~(\ref{fp_eq}) and 
(\ref{eq_23}) rests on the different assumptions of the underlying stochastic process. As it is well known, the standard diffusion operator  in  Eq.~(\ref{fp_eq}) assumes a Brownian random walk, whereas, in the context of the continuous time random walk (see for example Ref.~\cite{metzler} and references therein) the fractional derivative operators in 
Eq.~(\ref{eq_23}) assume an $\alpha$-stable Levy process describing the jumps of particles. 
One might naively expect that these two cases encompass all the fundamentally different 
transport operators  in the absence of memory and correlations. This expectation is based on the fact that 
according to the generalized central limit theorem the sum of individual random particle displacements asymptotically converge to either a Gaussian (described by the standard diffusion equation) or an $\alpha$-stable L\'evy distribution (described by the $\alpha$-stable fractional diffusion equation). However, an issue of significant physical relevance is that the convergence to the Gaussian or $\alpha$-stable behaviors could be extremely slow in problems of practical interest. 
For example, in the case of truncated L\'evy flights, it is known \cite{feller,mantegna,shlesinger} that the 
crossover time scale, $\tau_c$, of the convergence to Gaussian behavior scales algebraically $\tau_c \sim \lambda^{-\alpha}$ where $\lambda$ is the level of truncation.   Such slow algebraic convergence indicates that for the intermediate time scales of practical interest, these type of  processes cannot accurately be described as  Gaussian  or $\alpha$-stable.  
Motivated by this, Ref.~\cite{cartea_del_castillo_2007} considered the continuous time random walk for jump processes corresponding to  general L\'evy processes in the L\'evy-Khintchine representation, and obtained integro-differential operators describing transport in the continuous, fluid limit. These operators contain as special cases the regular (Laplacian) and the $\alpha$-stable fractional diffusion operators (Riemann-Liouville fractional derivatives), and,  most importantly, allow the incorporation of a richer class of stochastic processes of physical relevance.  In particular, for the case of truncated L\'evy processes, 
Ref.~\cite{cartea_del_castillo_2007} proposed the truncated fractional diffusion transport operator
\bq
\label{lambda_operator}
\partial_x^{\alpha,\lambda} P = {\cal D}_x^{\alpha,\lambda} P + v \partial_x P -\nu P \, ,
\eq
where ${\cal D}_x^{\alpha,\lambda}$ is the $\lambda$-truncated fractional derivative operator of order $\alpha$,
defined as 
\begin{equation}
\label{eq_28}
{\cal D}_x^{\alpha,\lambda} = l e^{-\lambda x}\,
_{-\infty}D_x^\alpha\, e^{\lambda x}\, +r e^{\lambda
x}\, _{x}D_{\infty}^\alpha\, e^{-\lambda x}  \, ,
\end{equation}
where $_{-\infty}D_x^\alpha$ and $_{x}D_{\infty}^\alpha$ are the Riemann-Liouville derivatives in Eq.~(\ref{fd_rl}).
The effective advection velocity is defined as
\begin{eqnarray}
\label{v_lambda}
v=\left\{
\begin{array}{rc}
0    \qquad 0<\alpha<1
\\
\frac{\chi \alpha \theta \lambda^{\alpha-1}}{\left| \cos \left( \alpha \pi/2\right) \right|} \qquad 1<\alpha<2 
\end{array}
\right.
\end{eqnarray}
and
\begin{equation}
\label{eq_29}
  \nu = -\frac{ \lambda^\alpha}{\cos\left(
\alpha \pi /2\right)}\, .
\eq
As expected, for $\lambda=0$, the transport operator in Eq.~(\ref{lambda_operator}) reduces to the $\alpha$-stable fractional diffusion operator in Eq.~(\ref{alpha_operator}). 
Based on the previous discussion, we propose the following spatially-tempered Fractional Fokker-Planck equation
\bq
\label{tempered_ffp}
\partial_t P = \partial_x \left[ P \partial_x V \right] + \chi  \partial_x^{\alpha, \lambda} P \, ,
\eq
to study the role of truncation effects on L\'evy flights in the presence of external potentials in the intermediate asymptotic regime. In particular, the study presented here of the role of truncation in fluctuation-driven transport in the presence of L\'evy flights is based on the numerical integration of Eq.~(\ref{tempered_ffp}).
Taking the  Fourier transform of Eq.~(\ref{tempered_ffp}), we get
\bq
\label{ft_eq}
\partial_t \tilde{P}= \Lambda(k) \tilde{P} - i k \,  {\cal F} \left[ P \partial_x V \right](k) \, ,
\eq
where $\Lambda(k)$ is the logarithm of the characteristic exponent of the tempered L\'evy process 
\bq
\begin{array}{l}
\label{eq_26}
\Lambda=
\frac{- \chi}{2 \cos(\alpha\pi/2)}  \times
\\ \\
\left\{
\begin{array}{ll}
(1+\theta) (\lambda +ik )^{\alpha }+
  (1-\theta) (\lambda -ik)^{\alpha }
- 2\lambda ^{\alpha }
\mbox{,} \\ \\
(1+\theta)(\lambda +ik )^{\alpha }+
(1-\theta)(\lambda -ik )^{\alpha }-
2 \lambda^{\alpha }- 2 i k\alpha \theta \lambda ^{\alpha -1}  \, ,
\end{array}
\right.
\end{array}
\eq
for $0<\alpha<1$ and $1<\alpha\leq2$ respectively. 

\section{Numerical results: ratchet current}

The spatio-temporal evolution of the pdf in the finite domain $(0,1)$ was obtained by solving numerically the spatially tempered fractional Fokker-Planck equation using a finite difference method. Following Ref.~\cite{del_castillo_2006}, we 
use regularized (in the Caputo sense)  fractional derivatives in space. 
However, here we factorize the regularized operators as $_0^cD_x^\alpha=_0^cD_x^{-(2-\alpha)} \partial^2_x$, and
$_x^cD_1^\alpha=_x^cD_1^{-(2-\alpha)} \partial^2_x$,  
and discretize the 
 fractional integral operators,  $_0^cD_x^{-(2-\alpha)}$ and  $_x^cD_1^{-(2-\alpha)}$,  using the Grunwald-Letnikov representation. The 
second order derivative, $\partial^2_x$, was discretized 
using  a central difference method. 
Following  Ref.~\cite{dcn_physA}, we discretized the  time evolution using an operation splitting method
that separates the fractional derivative operators from the advection and potential terms.  The 
fractional derivative operators (half) time step was done using a weighted average Crank-Nicholson method.

Figure \ref{dynamics} shows snapshots in time of the numerically calculated pdf's for different levels of truncation.  
The initial condition corresponds to a distribution of particles  localized in the potential well at the middle of the computational domain
\bq
P(x,0)=\frac{1}{\sigma \sqrt{2 \pi}} \exp \left[ -\frac{\left(x-1/2 \right)^2}{\sigma^2} \right] \, ,
\eq
with $\sigma=0.12$. 
Compared to the $\lambda=0$ ($\alpha$-stable) case, it is observed that the tempering reduces the ``leakage" of the pdf out of the potential well. However, for both $\lambda=0$ and $\lambda=3$ the profile peaks are higher on the right than on the left, indicating the presence  of a net current. The existence of the current for $\lambda=0$ is consistent with the results previously reported in Ref.~\cite{dcn_physA}. The understanding and characterization of the current for $\lambda \neq 0$ is one the objectives of this section. 
\begin{figure}
\includegraphics[scale=0.4]{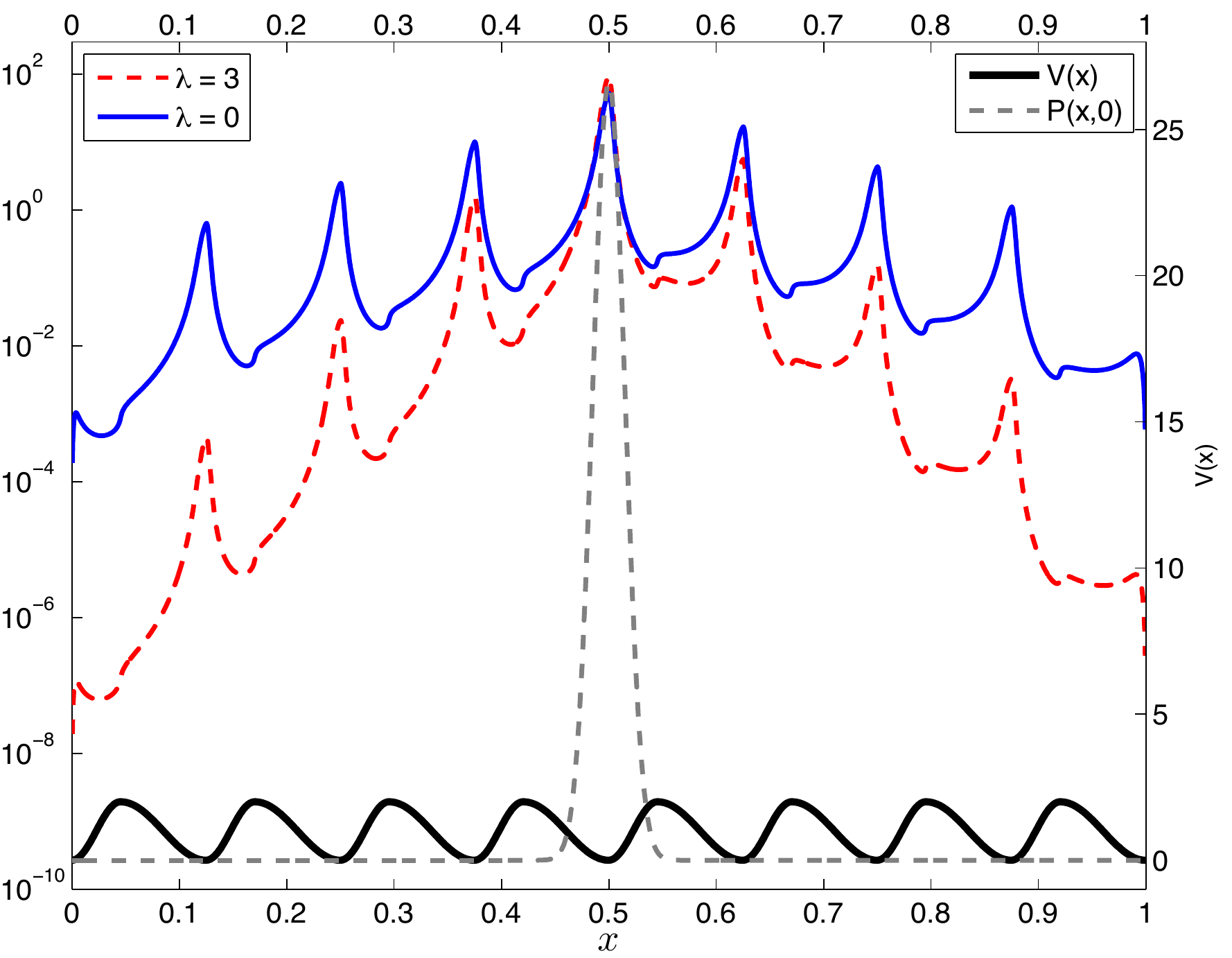} 
\caption{
\label{dynamics} 
Time dependent solutions of the truncated factional Fokker-Planck equation (\ref{tempered_ffp}) for
 $\lambda=0$ and $\lambda=3$,  with $\alpha=1.25$, $\theta=0$, $V_0=1$, and $\chi=0.5$. The solid line at the bottom is the potential in linear-linear scale. The pdfs for $\lambda=0$ and $\lambda=3$ are sown in a log-linear scale. 
}
\end{figure}

Given a solution, $P$,  of the tempered fractional Fokker-Planck equation, the current is defined as the rate of change of the first moment, 
\bq
\label{current}
J = \frac{d \langle x \rangle}{dt}=-i \frac{\partial }{\partial t} \left( \frac{\partial \tilde{P}}{  \partial k} \right)_{k=0} \, .
\eq
From Eqs.~(\ref{current}) and (\ref{ft_eq}), it follows that
\bq
\label{ft_current}
J=- {\cal F} \left[ P \partial_x V \right]_{k=0}-i \left(\frac{d \Lambda}{dk}\right)_{k=0}
\eq
In the $\alpha$-stable case, 
\begin{eqnarray}
\lambda=0 \, \Longrightarrow \qquad 
\left(\frac{d \Lambda}{dk}\right)_{k=0}=  \left\{
\begin{array}{rc}
\infty \,  \qquad 0 < \alpha < 1 \\
0 \,  \qquad 1< \alpha < 2 \, .
\end{array}
\right.
\end{eqnarray}
According to this, without tempering, the current is not well-defined for $\alpha<1$ due to the divergence of the first moment of L\'evy distributions. However, the introduction of tempering leads to a well-defined current because in this case 
 \begin{eqnarray}
 \lambda \neq 0 \, \Longrightarrow  \qquad 
\left(\frac{d \Lambda}{dk}\right)_{k=0}=  \left\{
\begin{array}{rc}
-\frac{\chi \alpha \theta i \lambda^{\alpha-1}}{\left| \cos \left(\alpha \pi /2 \right) \right|}\, \qquad 0 < \alpha < 1 \\
0 \, \qquad  1< \alpha < 2 \, .
\end{array}
\right.
\end{eqnarray}

The asymptotic steady state current  can be obtained numerically by advancing the solution for $P$ to long times as done in Ref.~\cite{dcn_physA}. However, here we follow a more direct approach based on the computation of the steady state solution in Fourier space, which can be formulated as an eigenvalue problem as shown below. 
For a general periodic potential, $V(x+L)=V(x)$, 
\bq
\partial_x V=\sum_{n=-\infty}^{\infty} c_n e^{i k_n x} \, ,
\eq
where $k_n=2 \pi n /L$, we have
\bq
{\cal F} \left[ P \partial_x V \right]= \sum_{n=-\infty}^{\infty} c_n
 \tilde{P} \left(k+k_n, t \right) \, .
 \eq
 Evaluating this expression at $k=0$ and substituting into Eq.~(\ref{ft_current}) we get the following expression for the current
 \bq
 \label{Current_Fourier}
  J= - \sum_{n=-\infty}^{\infty} c_n \tilde{P}_n -i \left(\frac{d \Lambda}{dk}\right)_{k=0} \, ,
 \eq
 where $\tilde{P}_n= \tilde{P}\left(k=k_n \right)$  is the Fourier transform of the steady state solution.
To determine the time evolution of the $n$-th Fourier mode, $\tilde{P}_n(t)$, we evaluate Eq.~(\ref{ft_eq}) at $k=k_n$  and get  the following set of coupled differential equations
 \begin{equation}
 \label{diff}
\partial_t \tilde{P}_{n} = \Lambda_{n} \tilde{P}_{n} - i k_n \sum_{j = -\infty}^{\infty} c_{j} \tilde{P}_{n+j}  \, ,
\end{equation}
where $\Lambda_n=\Lambda(k=k_n)$.
The steady state asymptotic current, $J(\infty)$, is obtained by setting $\partial_t \tilde{P}_{n}=0$ in Eq.~(\ref{diff}) and solving the set of algebraic equations
 \begin{equation}
 \label{alg}
\Lambda_{n} \tilde{P}_{n} - i k_n \sum_{j = -\infty}^{\infty} c_{j} \tilde{P}_{n+j} =0 \, .
\end{equation}
For the numerical solution of Eqs.~(\ref{diff}) and (\ref{alg}), we truncated the infinite series at a large value of $n=N$
and solved the system with the normalization constrain  $\tilde{P}_{0}=1$ (as required by the conservation of probability). 
In the calculations reported here $N$ ranged from $500$ to $2000$. The unique solution $\{\tilde{P}_n \}$ was then used in 
Eq.~(\ref{Current_Fourier}) to evaluate the current. The corresponding reconstructed pdf in real space gives the reduced pdf, $P_r$, 
\bq
P_{r}(x)=\sum_{n=-N}^{N} P_n e^{i k_n x} \, ,
\eq
which is the steady state solution of the tempered fractional Fokker-Planck equation in a periodic domain. This solution can also be obtained by ``folding" the unbounded domain solution in the $(0,L)$ domain
\cite{reimann}
\begin{equation}
P_{r}=\sum_{n=-\infty}^{\infty} P(x+nL,t)  \, .
\end{equation}

Figure~\ref{steady_alpha} shows the reduced steady state pdf in the $\alpha$-stable case, i.e., $\lambda=0$, for different values of  $\alpha$. 
In all the numerical results, the current has been non-dimensionalized using the scale $L/\tau$, where $\tau=L^\alpha/\chi$ is the fractional diffusion time scale.  As expected, for $\alpha=2$, $P_r$ corresponds to the Boltzmann distribution $P_r(x)=(1/Z) \exp \left( -V(x)/\chi \right)$, where $Z$ is the normalization constant. As the value of $\alpha$ decreases, the pdf significantly departs from the Boltzmann distribution and develops the asymmetry responsible for the finite net current observed for $\alpha<2$. 
The dependence of the reduced pdf on $\lambda$ for a fixed value of $\alpha$ is shown in Fig.~\ref{steady_lambda}. 
The leakage of the pdf outside the potential wells is significantly reduced for large values of $\lambda$. As we will discuss later, in general the current decreases with increasing $\lambda$, but the pdf does not approach the diffusive Boltzmann distribution. 
\begin{figure}
\includegraphics[scale=0.4]
{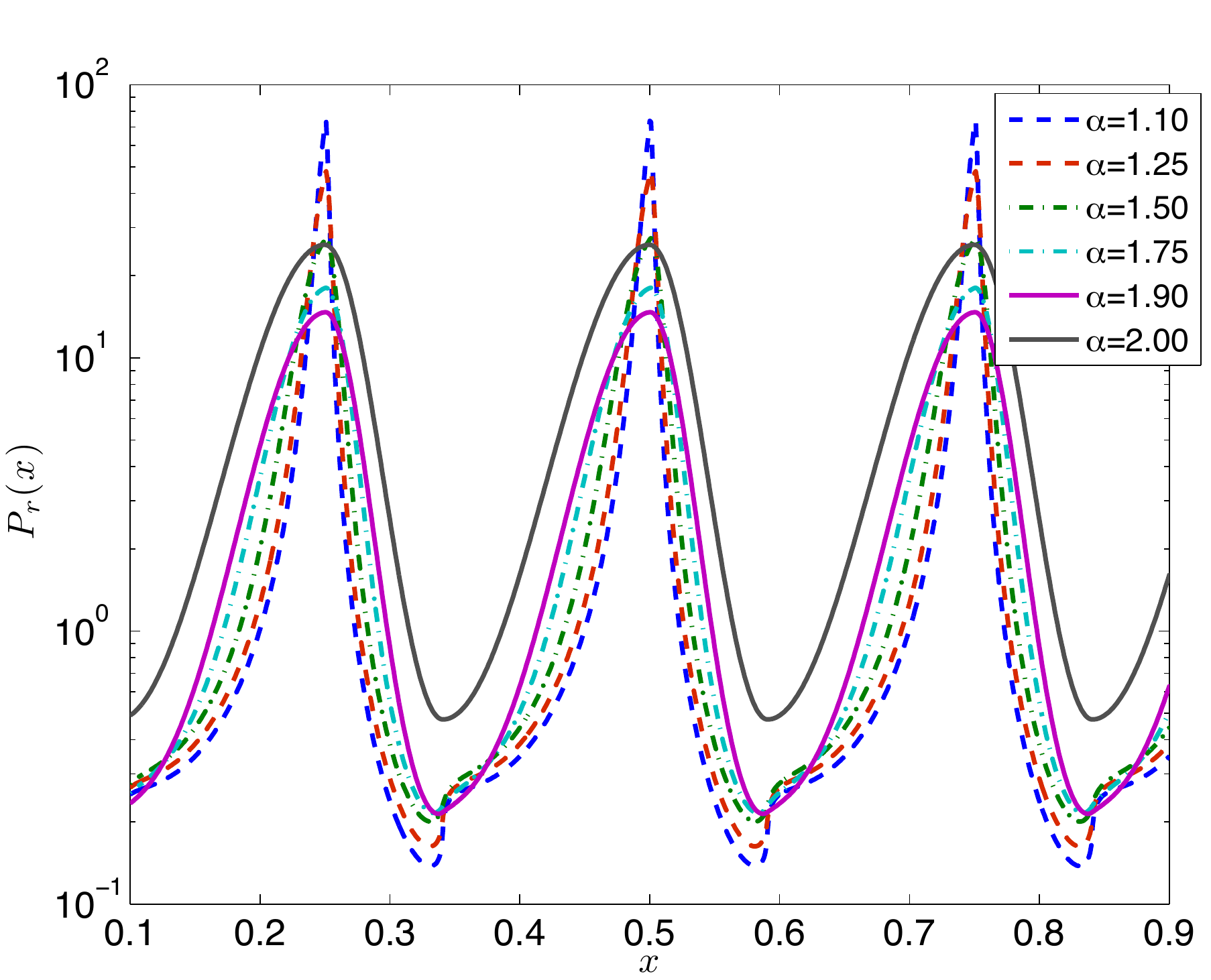} 
\caption{
\label{steady_alpha} 
Steady state solution of the reduced pdf in the periodic domain for  different values of $\alpha$, and $\lambda=0$, $\theta=0$, $V_0=1$,  
$A=-0.274$, and $\chi=0.5$.}
\end{figure}
\begin{figure}
\includegraphics[scale=0.4]
{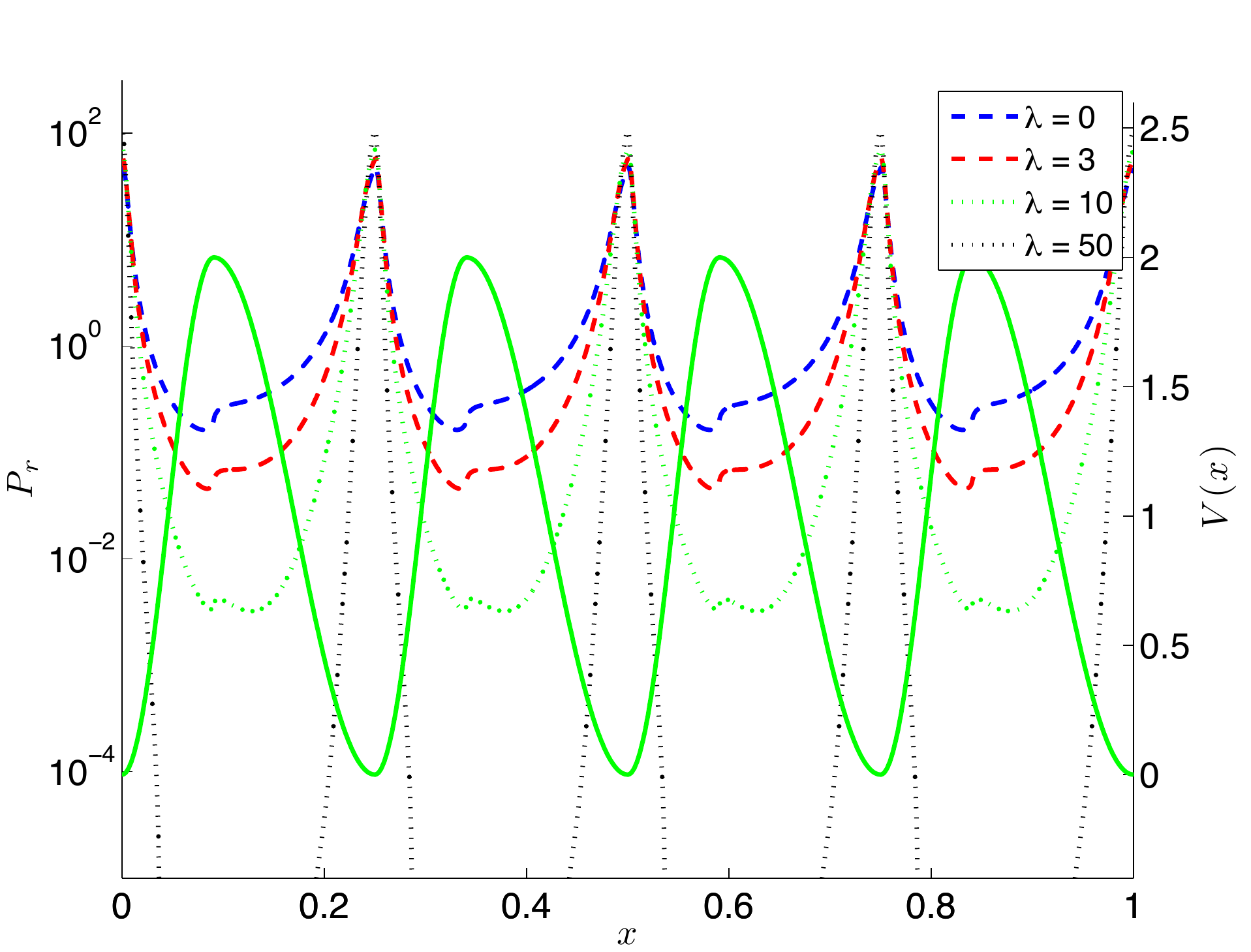} 
\caption{
\label{steady_lambda}  
Steady state solution of the reduced pdf in the periodic domain for  different values of $\lambda$, and  $\alpha=1.25$, $\theta=0$, $V_0=1$, 
$A=-0.274$ and $\chi=0.5$. The solid line is the ratchet potential.}
\end{figure}
Figure~\ref{J_vs_alpha} shows the dependence of the current on the value of $\alpha$ for different levels of  tempering. 
 In agreement with the results reported in Ref.~\cite{dcn_physA}, in the absence of tempering, $\lambda=0$, a net current is observed for $\alpha <2$ with a maximum around $\alpha \approx 1.4$. Most importantly, as the figure shows, a finite (albeit smaller) current remains for $\lambda \neq 0$. That is, although for $\lambda \neq 0$, the statistics of the tempered fractional operator in Eq.~(\ref{lambda_operator}) ultimately converges to Gaussian behavior \cite{cartea_del_castillo_2007}, in the presence of  a  ratchet potential a net non-equilibrium current persists in the steady state asymptotic regime. As expected, regardless of the value of $\lambda$, in the limit $\alpha=2$, the noise becomes Gaussian and the current vanishes.
 It is observed that as the truncation increases, the value of $\alpha$ for which the maximum current is attained shifts to the right, and for large values of $\lambda$ the strength of the current exhibits only a weak dependence on $\alpha$. 
 To study the rate of convergence to the steady state asymptotic current, we show  in Fig.~\ref{delta_J_time}  $\delta J(t)= J(t)-J(\infty)$ as function of time where $J(t)$ is the time dependent transient current and $J(\infty)$ is the steady state asymptotic current. An exponential convergence of the form 
 \bq
 \label{deltaJ}
 \delta J  \sim e^{-\eta t} \, ,
 \eq
 is observed, and as Table~I shows, the decay rate, $\eta$, exhibits a very weak dependence on the value of $\alpha$ and $\lambda$. 

\begin{figure}
\includegraphics[scale=0.4]
{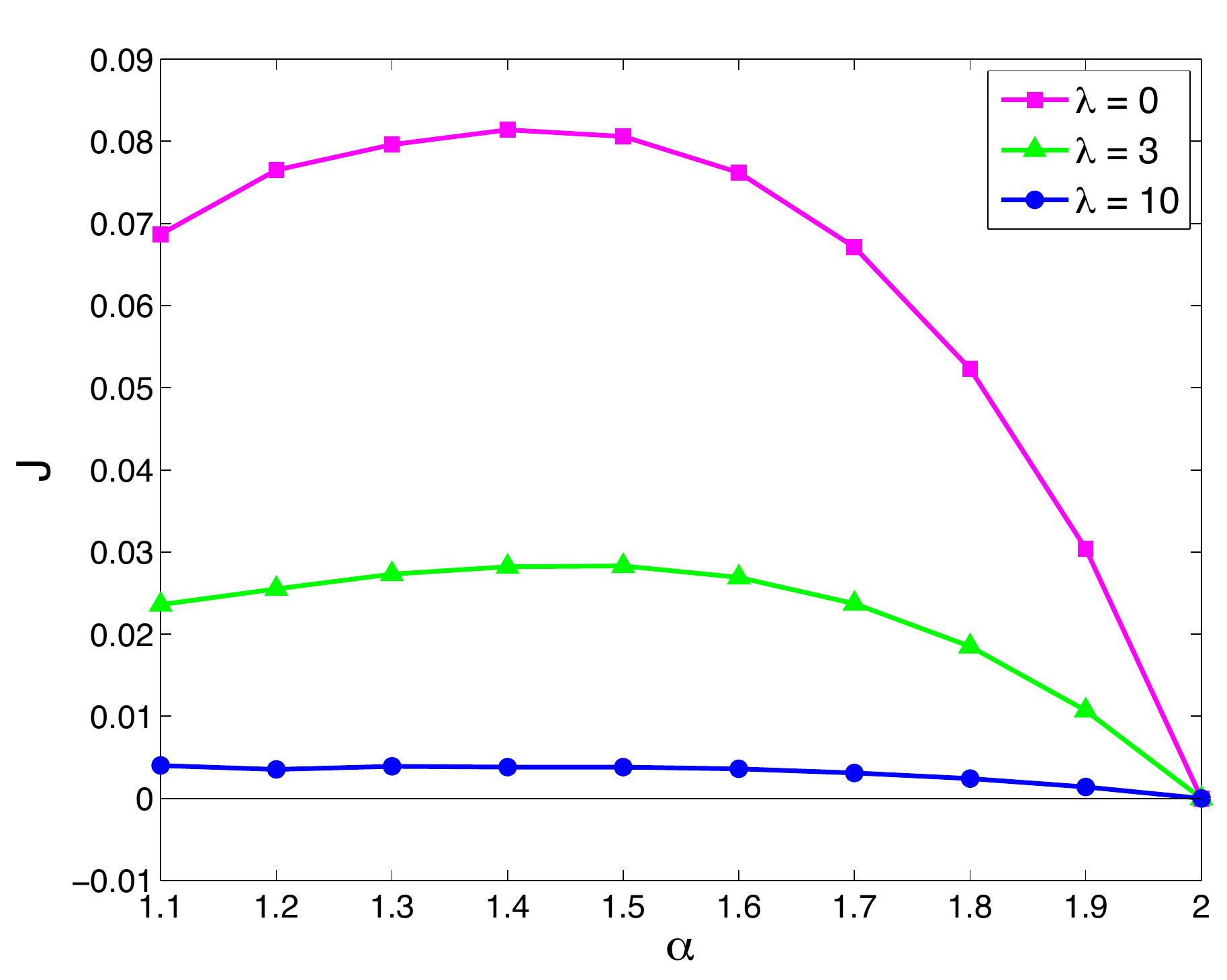} 
\caption{
\label{J_vs_alpha} 
Ratchet current, $J$, as function of $\alpha$, for different levels of truncation, $\lambda$, with
$\theta=0$, $A=-0.6$, $V_0=1$, and $\chi=0.5$.
}
\end{figure}
\begin{figure}
\includegraphics[scale=0.40]
{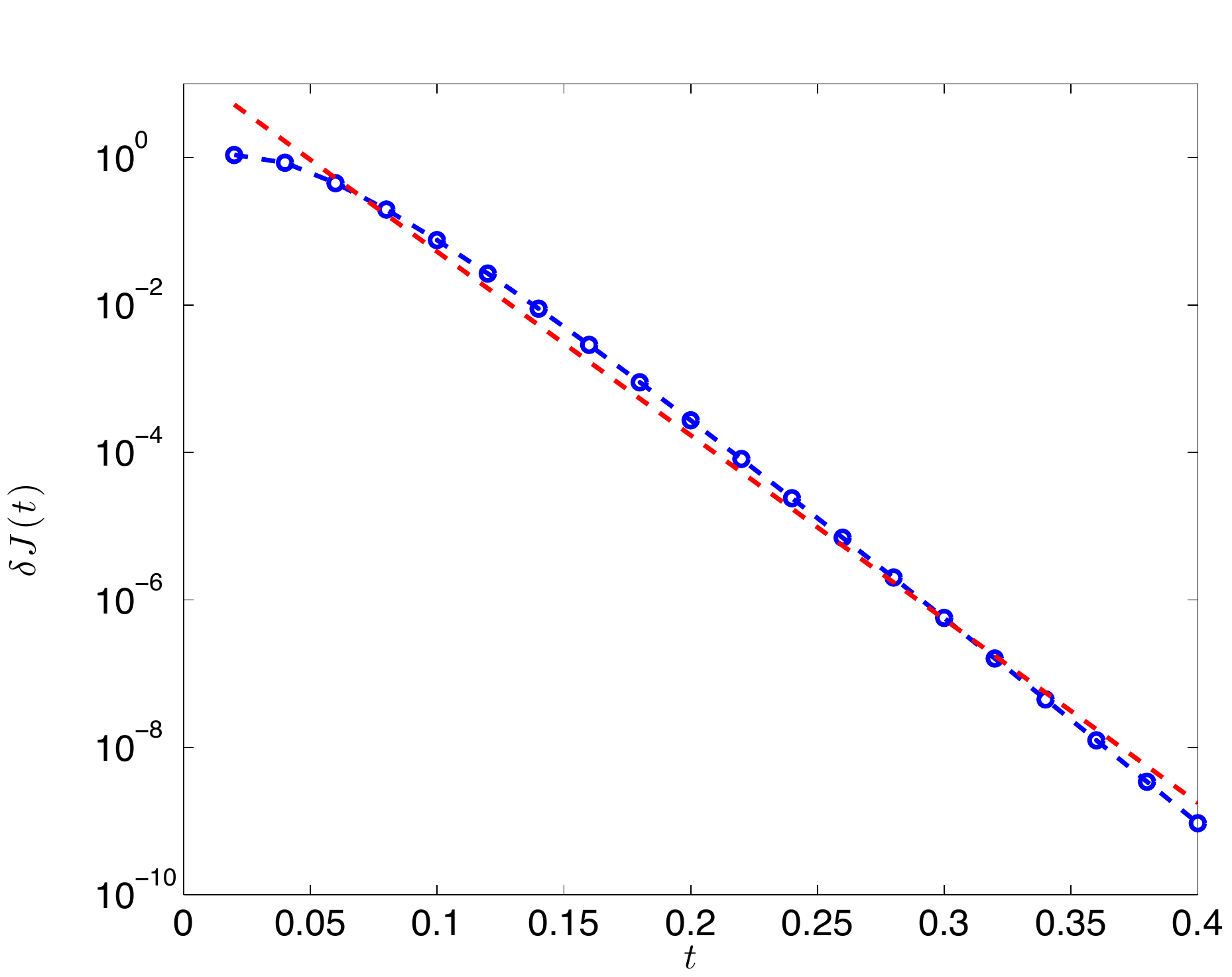} 
\caption{
\label{delta_J_time} 
Exponential convergence of ratchet current as function of time. The plot shows $\delta J=J(\infty)-J(t)$ where $J(\infty)$  is the steady state asymptotic current (obtained from the solution of Eq.~(\ref{alg})) and $J(t)$ is the time dependent transient current (obtained from the solution of Eq.~(\ref{diff}). 
The parameter values are $\alpha=1.5$, $\lambda=10$, 
$\theta=0$, $A=-0.274$, $V_0=1$, and $\chi=0.5$. The dashed line is an exponential fit of the form $\delta J \sim e^{-\eta t}$
with $\eta ~\sim 57$. Table~I shows the exponential decay rate for different values of $\alpha$ and $\lambda$. 
}
\end{figure}


\begin{table}[htdp]
\caption{Dependence of exponential decay rate, $\eta$, of transient current in Eq.~(\ref{deltaJ}) as function of $\alpha$ and $\lambda$ for $\theta=0$, $A=-0.274$, $V_0=1$, and $\chi=0.5$.  Case $\alpha=1.5$, $\lambda=10$ is shown in 
Fig.~\ref{delta_J_time}.} 
\begin{center}
\begin{tabular}{|c|c|c|c|c|c|c|c|c|c|}
\hline
\hline
$\alpha$ & 1.25 & 1.25 & 1.25 & 1.5 & 1.5 & 1.5 & 1.75 & 1.75 & 1.75 \\
\hline
$\lambda$  & 0 & 10 & 30 & 0 & 10 & 30 & 0 & 10 & 30 \\
\hline
$\eta$  & 64 & 58 & 58 & 60 & 57 & 58 & 52 & 51 & 53  \\
\hline
\end{tabular}
\end{center}
\label{default}
\end{table}

As the previous calculations show, as $\lambda$ increases the current is reduced, and a problem of significant interest is to find the rate of decay of the current in the asymptotic limit $\lambda \rightarrow \infty$. 
As Figs.~\ref{J_large_lambda_loglog}-\ref{J_large_lambda_loglin_less_1} show, depending on the value of $\alpha$, two asymptotic regimes are observed. For $\alpha$ near the Gaussian value $\alpha=2$, the current exhibits an algebraic decay for large $\lambda$ of the form
\bq
J  \sim \lambda^{- \zeta}\, ,
\eq
where $\zeta \approx 2.5$ for $\alpha=1.9$ and $\zeta \approx 6$ for $\alpha=1.75$. This slow algebraic scaling breaks down as $\alpha$ is reduced from the Gaussian value. In particular, as Fig.~\ref{J_large_lambda_loglin}  shows, for $1<\alpha<1.5$ the current decays exponentially fast 
\bq
J  \sim e^{- \xi \lambda}\, ,
\eq
with $\xi \approx 0.40$. As 
Fig.~\ref{J_large_lambda_loglin_less_1} verifies, the same scaling is observed for $\alpha <1$.

\begin{figure}
\includegraphics[scale=0.4]
{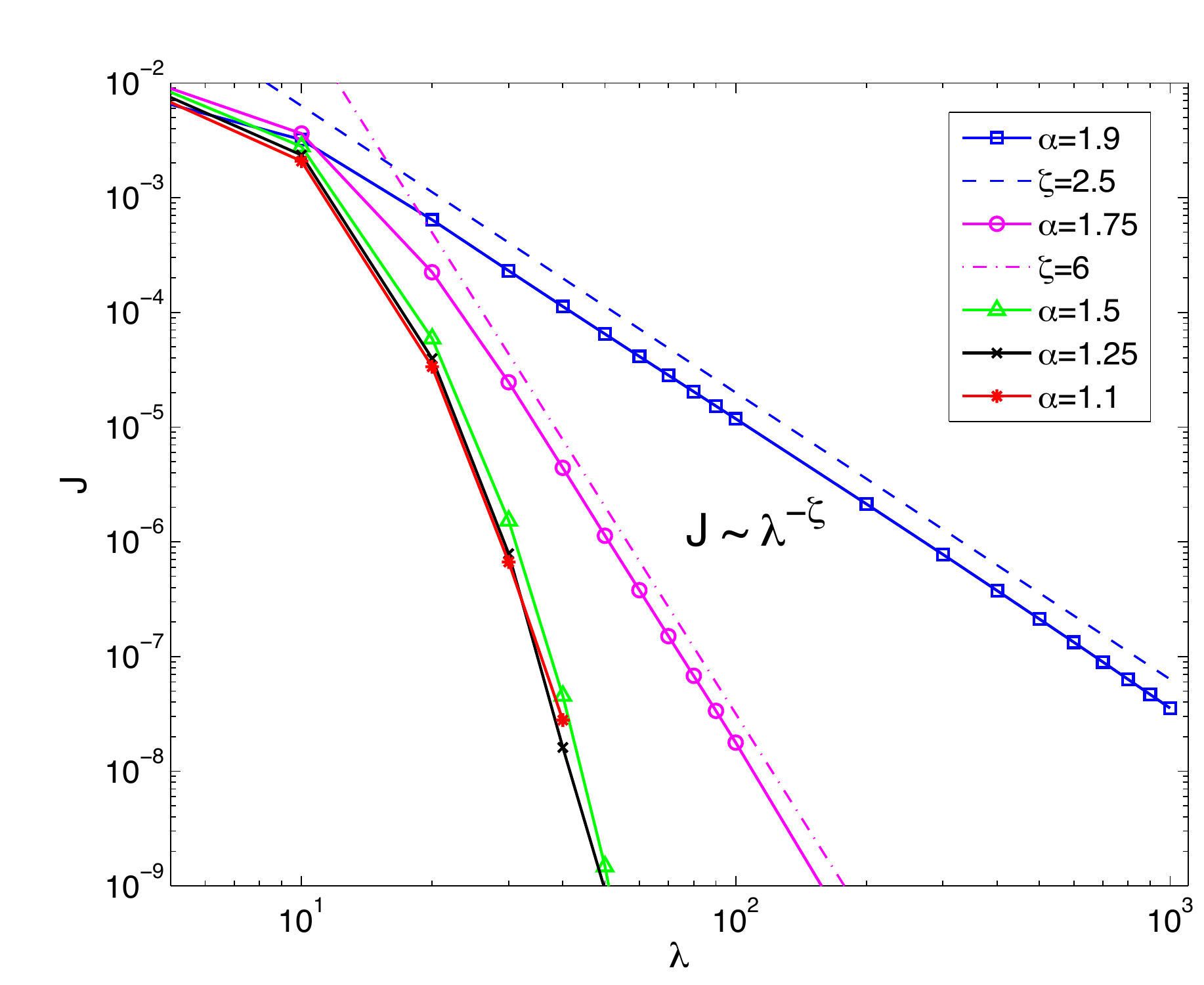} 
\caption{
\label{J_large_lambda_loglog}  
Decay of ratchet current, $J$, as function $\lambda$ for different values of 
$\alpha$ and $\theta=0$, $V_0=1$,  and $A=-0.274$. The Log-Log scale shows evidence of algebraic decay for $\alpha \geq 1.75$.
}
\end{figure}
\begin{figure}
\includegraphics[scale=0.4]
{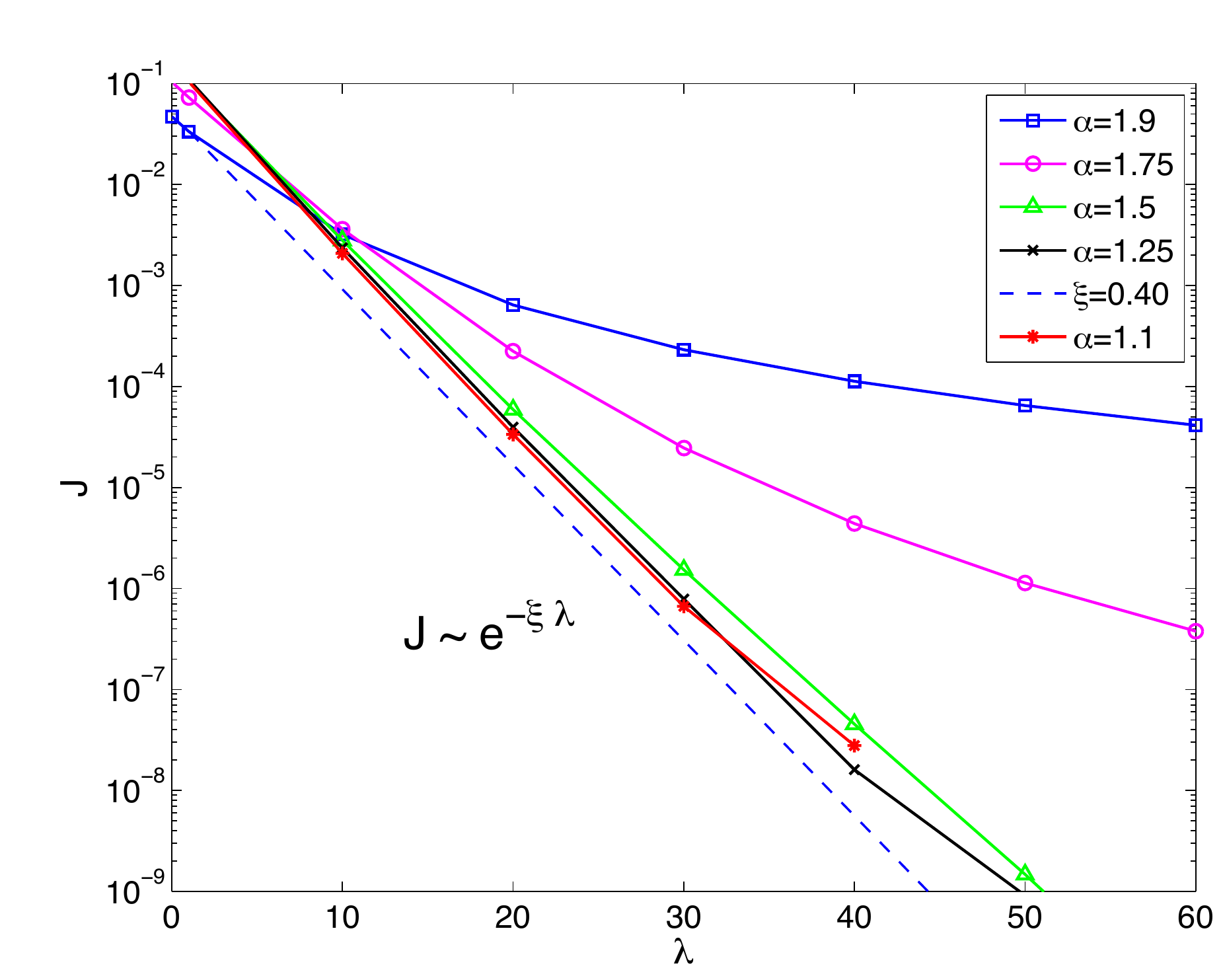} 
\caption{
\label{J_large_lambda_loglin}  
Decay of ratchet current, $J$, as function $\lambda$ for different values of 
$\alpha$ and $\theta=0$, $V_0=1$,  and $A=-0.274$. 
The Log-Linear scale shows evidence of exponential decay for $\alpha \leq 1.5 $.
}
\end{figure}
\begin{figure}
\includegraphics[scale=0.4]
{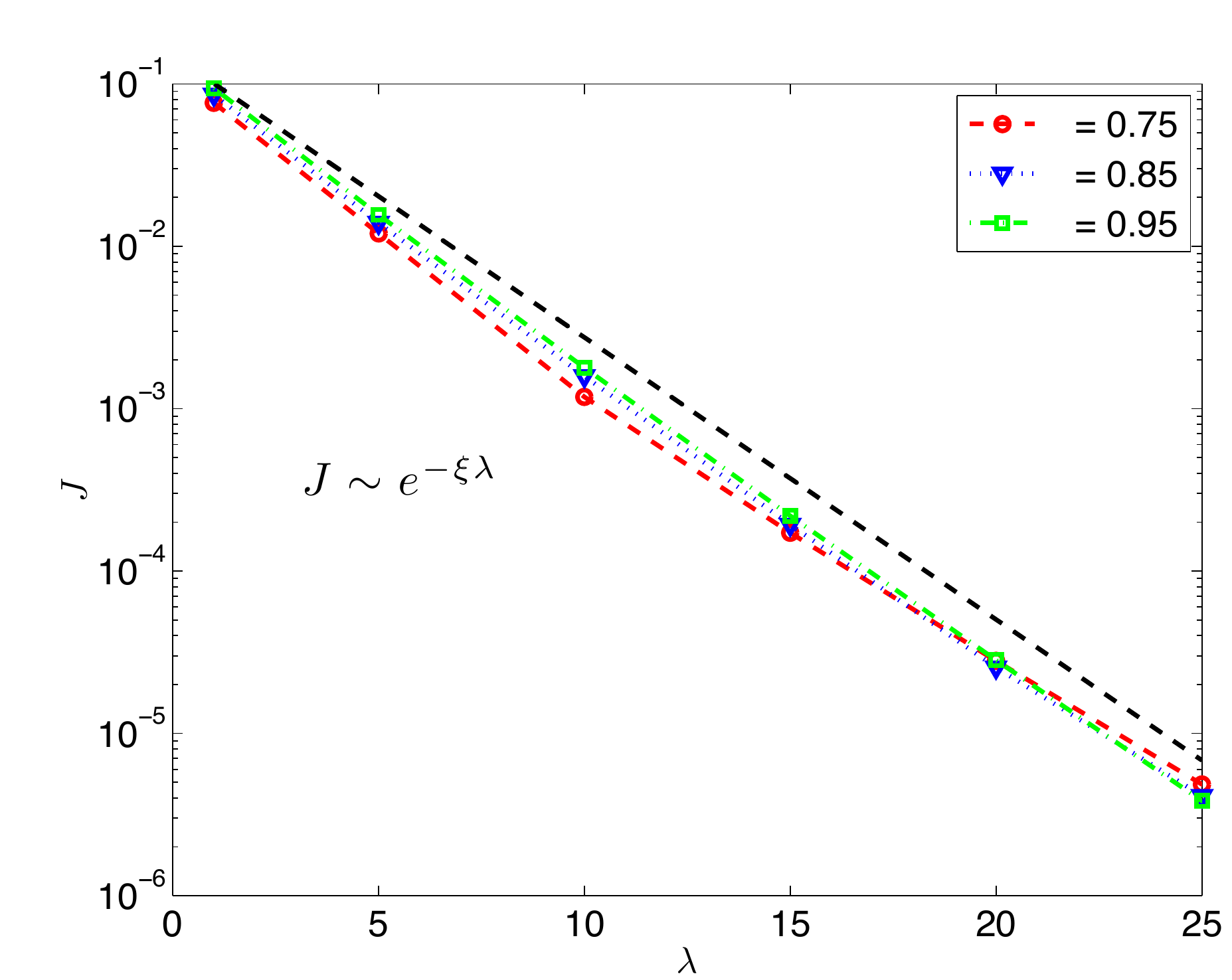} 
\caption{
\label{J_large_lambda_loglin_less_1}  
Same as Fig.~\ref{J_large_lambda_loglin} but for $\alpha <1$.
}
\end{figure}

Figure~\ref{J_vs_A} shows the dependence of the current, $J$, on the potential asymmetry, $A$. As expected, in all cases the current vanishes when the potential is symmetric, $A=0$.  Consistent with the results reported in \cite{dcn_physA}, the curve $\lambda=0$, which corresponds to an $\alpha$-stable L\'evy ratchet, shows the existence of a ratchet current in the direction of the steepest side of the potential. The main result of this study is that, even in the presence of truncation, a ratchet current is observed and its dependence on $A$ is qualitatively  similar to the $\alpha$-stable case. 
\begin{figure}
\includegraphics[scale=0.4]
{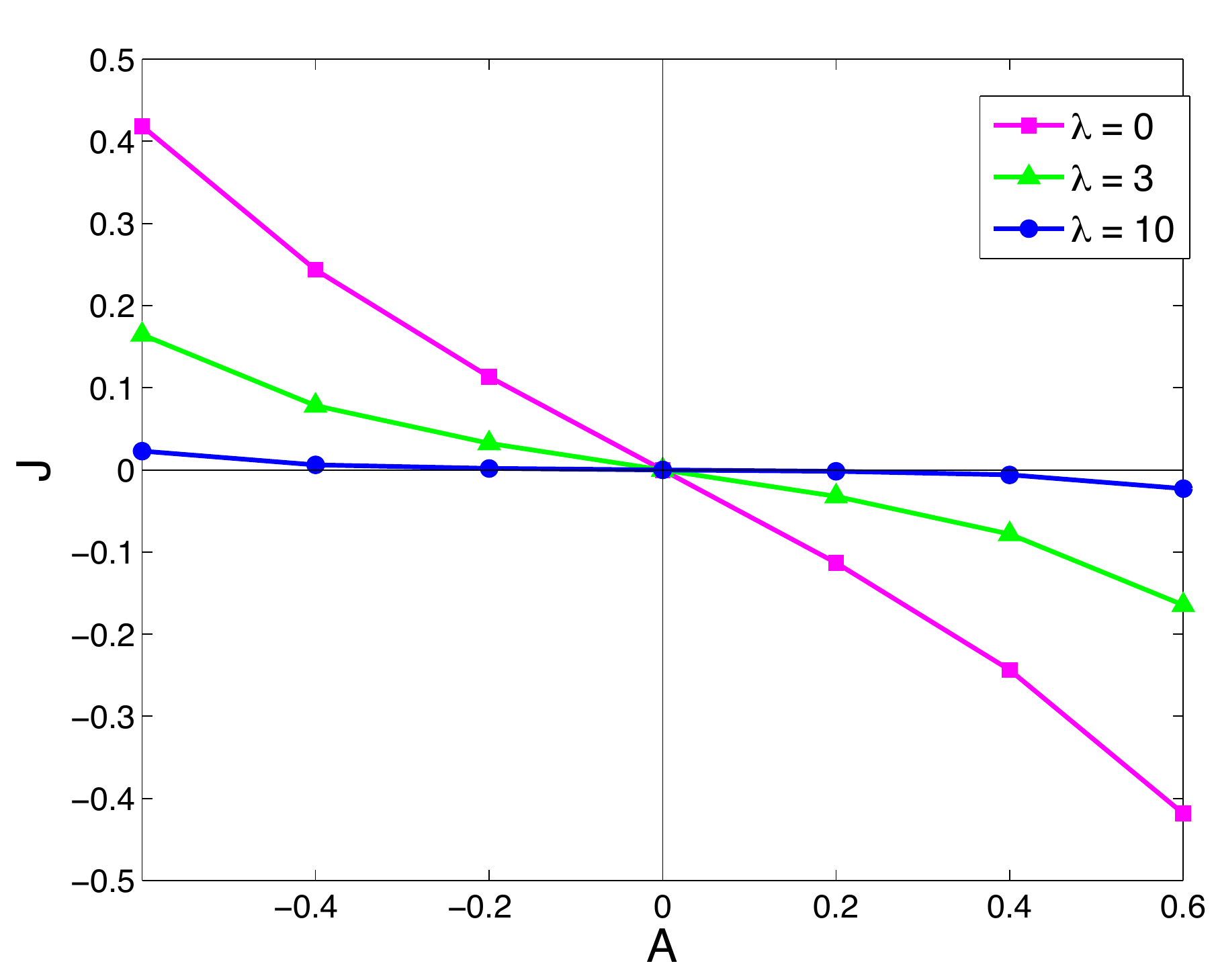} 
\caption{
\label{J_vs_A} 
Ratchet current, $J$, as function of ratchet potential asymmetry, $A$, for different levels of truncation, $\lambda$, $\alpha=1.5$, $\theta=0$, $V_0=1$, and $\chi=0.5$.
}
\end{figure}
To study the dependence of the current on an external tilting force, we computed the current  with an effective potential, $V_{eff}=V(x)-F x$ where $V(x)$ is the ratchet potential in Eq.~(\ref{potential}), and $F$ is a constant. 
As shown on Fig.~\ref{J_vs_F},  the $\lambda=0$ case recovers the $\alpha$-stable L\'evy ratchet results, and, as $\lambda$ increases, the strength of the current decreases. Note that the value of the stopping force, $F \approx -1$, i.e. the force  needed to cancel the ratchet current, seems to be independent of $\lambda$.  
\begin{figure}
\includegraphics[scale=0.4]
{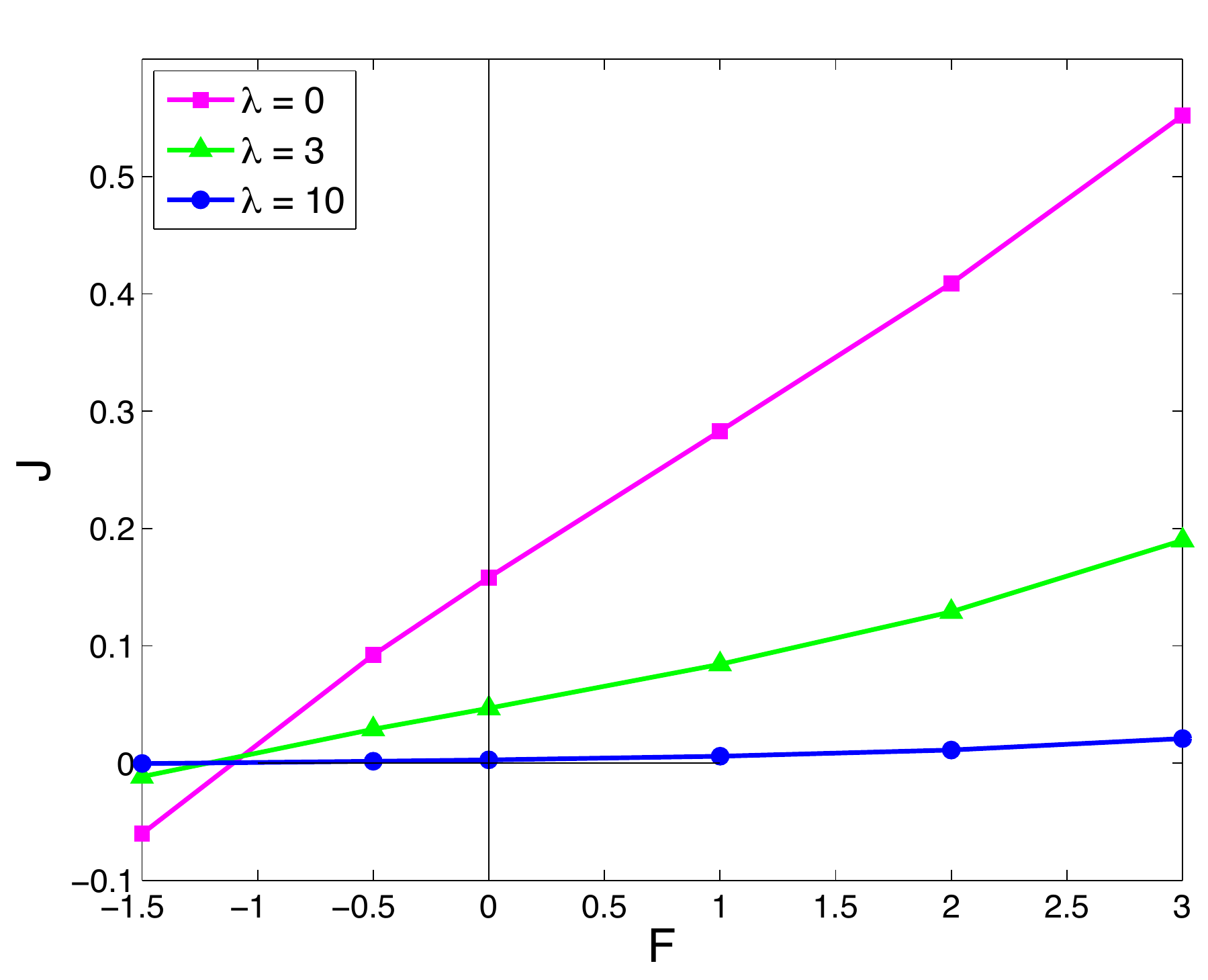} 
\caption{
\label{J_vs_F} 
Ratchet current, $J$, as function of external force, $F$, for different levels of truncation, $\lambda$,
$\alpha=1.5$, $\theta=0$, $V_0=1$, $A=-0.274$, and $\chi=0.5$.
}
\end{figure}

An interesting effect of the truncation is observed  when the L\'evy noise is asymmetric, i.e. when the weighting factors $l$, $r$ of the left and the right fractional derivatives in Eq.~(\ref{eq_28}) are different. 
\begin{figure}
\includegraphics[scale=0.4]
{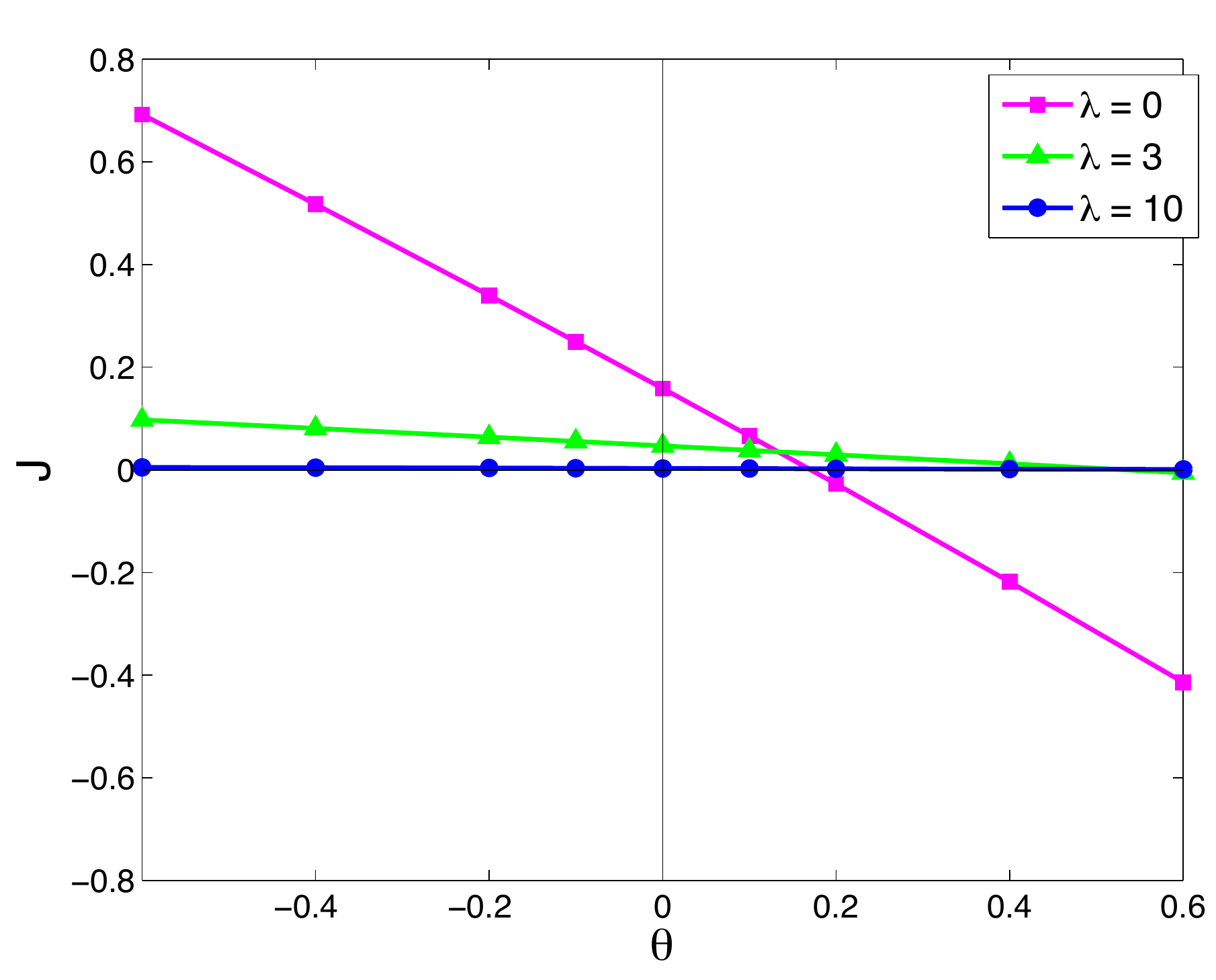} 
\caption{
\label{J_vs_theta} 
Ratchet current, $J$, as function of fractional diffusion asymmetry $\theta$ for different levels of truncation, $\lambda$, $\alpha=1.5$, $V_0=1$, $A=-0.274$,  and $\chi=0.5$.  
}
\end{figure}
As shown in Fig.~\ref{J_vs_theta}, in the absence of truncation, $\lambda=0$, the current can be reversed by biasing the asymmetry of the noise. In particular, for the case shown, increasing $\theta$ to a large enough value stops the positive current. This is consistent with the fact that for  $\theta>0$ the skewness of the L\'evy noise is negative and the probability of having long jumps to the right is reduced. As the value of   $\theta$ is further increased the probability of jumps to the left increases and this eventually leads to a negative current. 
A similar phenomenology is observed when the truncation is present, except that (in addition to the usual overall decrease of the magnitude of the current) the critical value of $\theta$ for current reversal increases. The fact that the critical $\theta$ for current reversal depends on $\lambda$ implies that for a range of $\theta$ values it is possible to reverse the current by changing $\lambda$ only. This interesting feature of truncated L\'evy ratchets is clearly illustrated in Fig.~\ref{J_vs_lambda}  where it is shown that for $\theta=0.275$ the truncation can in fact lead to a current reversal. 
\begin{figure}
\includegraphics[scale=0.4]
{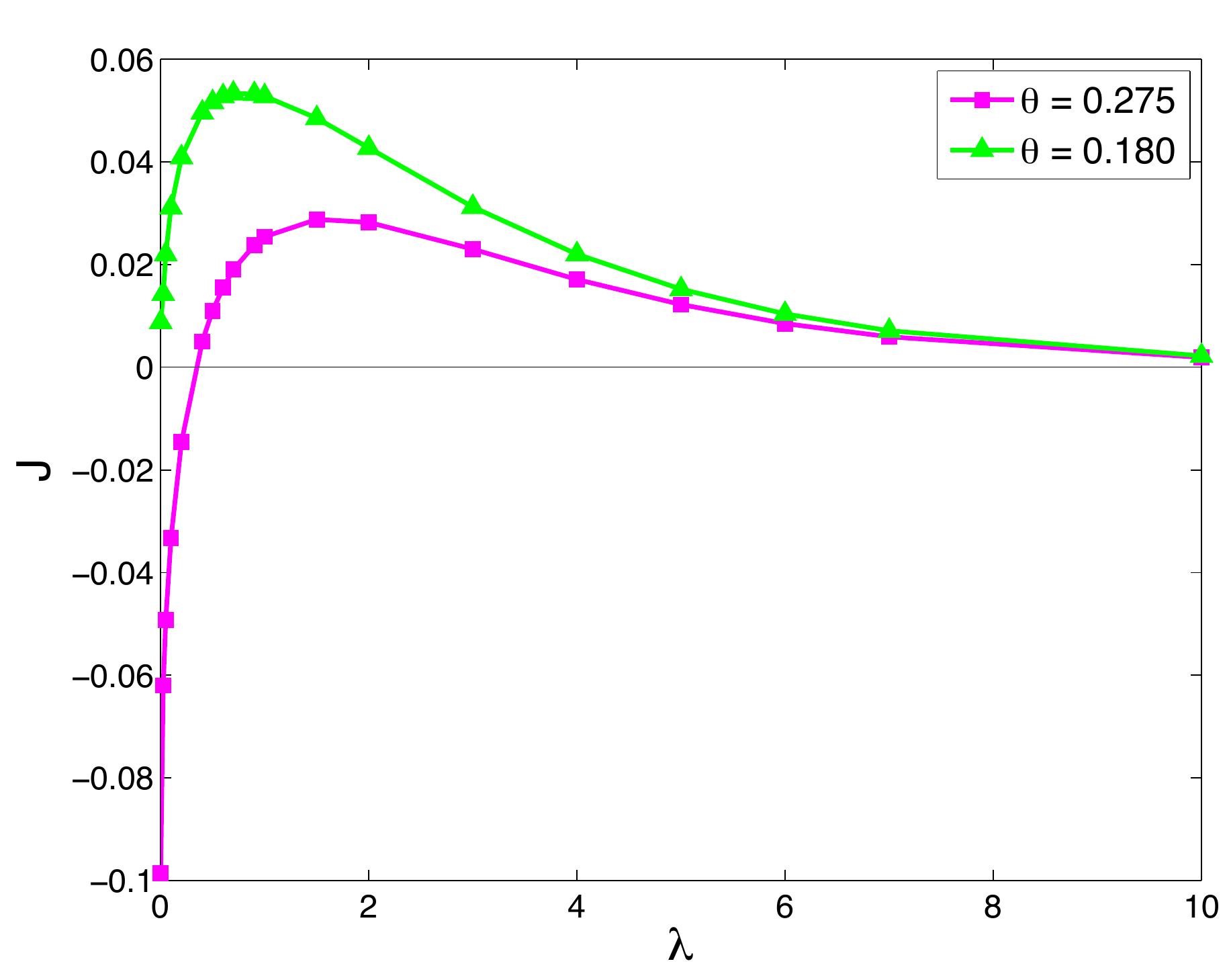} 
\caption{
\label{J_vs_lambda} 
Ratchet current, $J$, as function of truncation parameter $\lambda$, for different levels of fractional diffusion asymmetry $\theta$, $\alpha=1.5$,  $V_0=1$, $A=-0.274$, and $\chi=0.5$. For $\theta=0.275$, the current is reversed for $\lambda \approx 1/3$.
}
\end{figure}
This behavior is related to the fact that, as Eq.~(\ref{v_lambda}) shows, for $\theta \neq 0$ the truncation gives rise to  a finite advection term in the Fokker-Planck equation.  To conclude the numerical simulations of the steady state current, we show in  Fig.~\ref{J_vs_sigma}  the dependence of the current on  $\sigma=\chi^{1/\alpha}$, where $\chi$ is the diffusivity. In the absence of truncation the current is maximum for $\sigma \sim 1$. However, as the truncations increase the value of the diffusivity for which the maximum current is attained increases. 

\begin{figure}
\includegraphics[scale=0.4]
{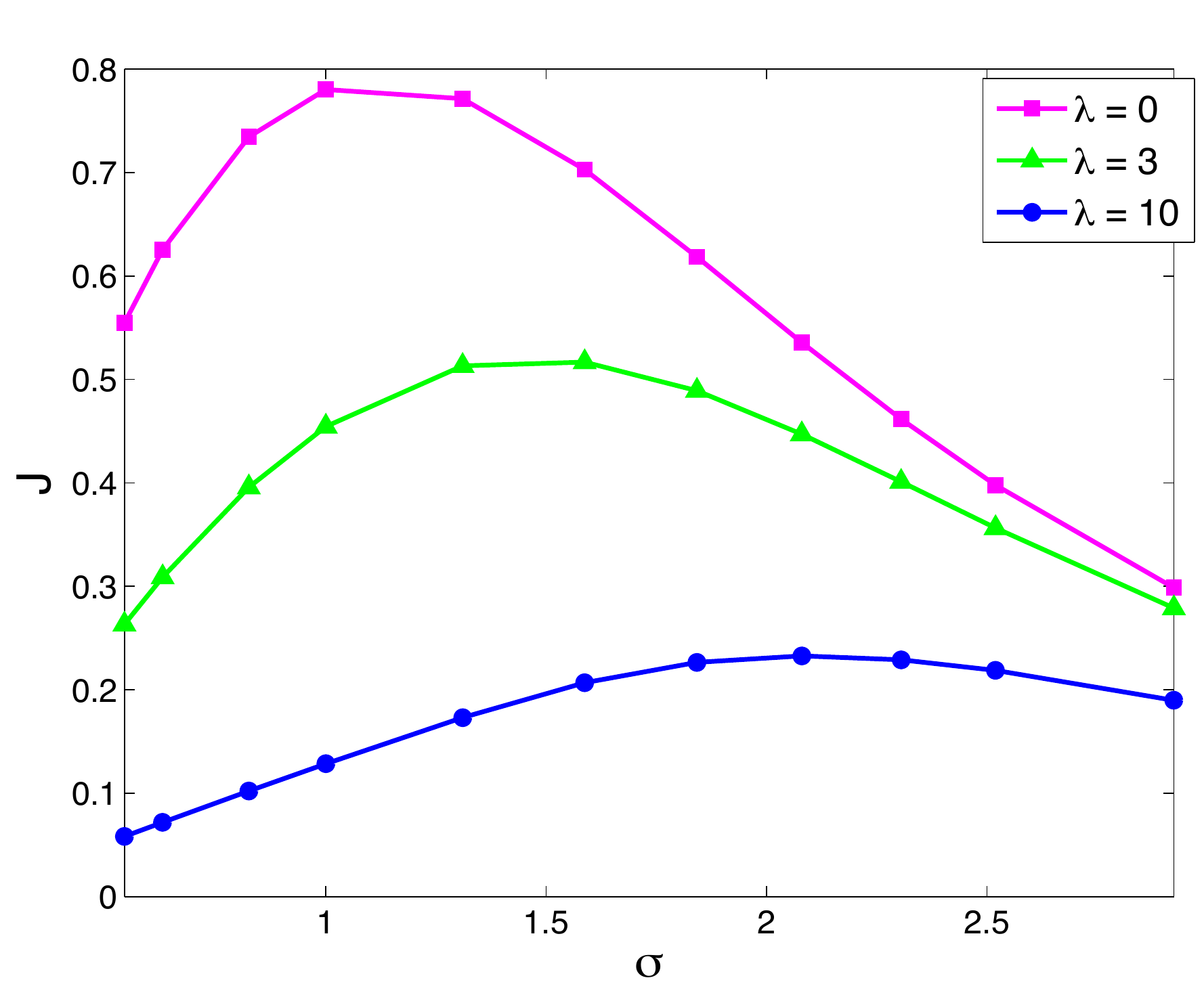} 
\caption{
\label{J_vs_sigma}  
Ratchet current, $J$, as function of $\sigma=\chi^{1/\alpha}$, for different levels of truncation, $\lambda$,
$\alpha=1.5$, $\theta=0$, $V_0=1$, and $A=-0.8$.
}
\end{figure}



\section{Conclusions}

The incorporation of tempering in models of anomalous transport due to L\'evy flights is motivated by the fact that in problems of practical interest, the moments of the pdfs represent physical quantities that cannot be unbounded. 
For example, although anomalously large displacements have been well documented in numerical simulations of turbulent transport and in fluid mechanics experiments, it is clear from the physical point of view that particle displacements cannot be arbitrarily large and that a cut-off or truncation has to be incorporated in the pdfs describing these processes. This was the intuition behind the original proposal of truncated L\'evy flight  in Ref.~\cite{mantegna} and the main motivation for the construction of spatially tempered fractional diffusion operators in Ref.~\cite{cartea_del_castillo_2007}.
In this paper we have studied the effects of spatial tempering on directed transport in L\'evy ratchets.  

Our study was based on the numerical solution of the time dependent and the steady state spatially tempered fractional Fokker-Planck equation. We used two complementary numerical methods. 
For the computation of the space-time evolution of the pdf in a finite size domain, we used a finite difference method based on the  Grunwald-Letnikov discretization of the truncated fractional derivatives. 
For the computation of the reduced pdf in a periodic domain as well as the time dependent and asymptotic steady state current, we used a Fourier based spectral method.
The main object of study was the dependence of the steady state current on the level of truncation, $\lambda$, the stability index $\alpha$, and the noise asymmetry $\theta$ as well as the asymmetry of the potential.

The general conclusion is that in most parameter regimes the inclusion of truncation does not change the qualitative dependencies of the current on the different system parameters; instead, the effect of truncation is mainly to decrease the magnitude of current.  However, an interesting exception to this rule occurs when the L\'evy noise is allowed to have a bias. In this case, truncation can lead to a reversal of current direction.
The persistence of ratchet currents for $\lambda \neq 0$ is interesting since the finite second moments of the truncated L\'evy statistics guarantee an eventual convergence to Gaussian statistics (in the absence of an external potential) and, and as it well-known, Gaussian fluctuations cannot create a net current in the absence of time dependent perturbations or an external tilting force. 
The current is observed to converge exponentially in time to the steady state value, and the decay rate exhibits only a weak dependence of $\alpha$ and $\lambda$.  
The decay of the current for increasing values of $\lambda$ exhibits different scaling properties depending on the value of $\alpha$. For large values of alpha, $1.75 \leq \alpha$, the current exhibits an algebraic decay and for  small values, $\alpha \leq 1.5$, the current exhibits an exponential decay. 

\subsection*{Acknowledgments}
The authors thank Prof. G. Morales for the valuable comments and suggestions to this work. 
This work has been supported by  the Oak Ridge
National Laboratory, managed by UT-Battelle, LLC, for the U.S.
Department of Energy under contract DE-AC05-00OR22725.
Work at UCLA is sponsored by NSF/DOE partnership grant SC0004663.

%


\end{document}